\documentclass[aps, prd, twocolumn, showpacs, nofootinbib]{revtex4}
\usepackage{amsmath}
\usepackage{graphicx}
\usepackage{dcolumn}
\usepackage{bm}
\usepackage{amssymb}
\usepackage{latexsym}
\usepackage{bm}
\usepackage{float}

\newcommand{\pa}{\partial}
\newcommand{\del}{\delta}
\newcommand{\eps}{\epsilon}

{\bf }

\newcommand{\half}{\frac{1}{2}}

\begin{document}


\title{Realization of cyons and anyons by  atoms }

\author{Jian Jing ${}^{a}$}
\email{jingjian@mail. buct. edu. cn}

\author{Yao-Yao Ma ${}^{a}$}

\author{Qing Wang ${}^{b}$}


\author{Zheng-Wen Long ${}^c$}

\author{Shi-Hai Dong ${}^d$}
\email{dongsh2@yahoo. com}

\affiliation{${}^a$ Department of Physics and Electronic, School of
Science, Beijing University of Chemical Technology, Beijing 100029,
P. R. China, }

\affiliation{${}^b$ College of Physics and Technology, Xinjiang University, Urumqi 830046, P. R. China, }

\affiliation{$^{c}$ Department of Physics, GuiZhou University, GuiYang, 550025, P. R. China}

\affiliation{{ $^{d}$ Laboratorio de Informaci\'on Cu\'antica, CIDETEC, Instituto Polit\'{e}cnico Nacional, UPALM, CDMX 07700, M{e}xico. }}

\begin{abstract}

We propose theoretical schemes  to realize the cyon and anyon by  atoms which  possess   non-vanishing   electric  dipole moments. To realize a cyon, besides the atom,  we  need a magnetic  field produced by a long magnetic-charged filament. To realize an anyon, however, apart from these we need a harmonic potential and an additional magnetic field produced by a uniformly distributed magnetic charges. We find that the atom will be an anyon when cooled down to the negligibly small kinetic energy limit. The relationship  between our results and the previous ones is investigated from the electromagnetic duality.

\end{abstract}

{\pacs {03. 65. Vf, 03. 65. Pm, 03. 65. Ge}}

\maketitle

\section{Introduction}

Anyons are quasi-particles with fractional spins. They only exist in the two-dimensional space since the rotation group in two-dimensional space is Abelian which cannot impose any constraint on the eigenvalues of the canonical  angular momentum \cite{anyons1, anyons2}. Anyons play important roles in understanding some two-dimensional phenomena \cite{hall, htc}.  It was shown that after coupling charged particles to the pure Chern-Simons gauge field on a plane, the eigenvalues of the canonical angular momentum of both non-relativistic and relativistic  charged  particle are fractional. This means that  anyons can be realized by coupling charged particles or charged fields to the pure Chern-Simons gauge field \cite{cs1, cs2, cs3, forte}. Recently, it was predicted theoretically that anyons may exist in the quantum Hall states \cite{zhang1,zhang2,zhang3}.

In Ref. \cite{zhang}, the author proposed a different approach to realize anyons, i.e., a trapped ion was coupled to two types of magnetic potentials instead of coupling  to  Chern-Simons gauge field. One is the dynamical magnetic potential whose magnetic field does not vanish in the area where the ion moves, the other is the Aharonov-Bohm type which is produced by an infinitely long-thin solenoid. As expected, the eigenvalues of the canonical angular momentum of the ion take quantized numbers in the unit of $\hbar$. However, the author found that the rotation property of the reduced model, which was obtained by cooling down the kinetic energy of the ion to the lowest level, behaved interestingly since the eigenvalues of the canonical angular momentum of this reduced model were fractional. The fractional part was proportional to the magnetic flux inside the solenoid. It is interesting to show that both of these two potentials  played important roles in producing fractional angular momentum although it seems that  only one of them appeared in the final result.

As  prototype of anyons, cyons also attracted much attention. A cyon is a compound system which was originally realized by  a  planar charged particle and a perpendicular infinitely long-thin solenoid  \cite{cyon1}. As expected,  eigenvalues of the canonical angular momentum  take integers. However, it is shown that the integer eigenvalues are divided into two parts: one is localized  at the charged particle which generally is  fractional and is related to the spin of the cyon, the other is also fractional which is located at the spatial infinity. It was pointed out that the part located at spatial infinity was irrelevant to  phenomena in a finite length scale and thus only the part attached to the charged particle was relevant since we are only interested in the behaviors of the charged particle \cite{cyon2}.
Therefore, the compound system has a fractional statistic property if two identical compound systems are interchanged.  Even though both  cyons and anyons were mostly realized by  charged particles before, one may wonder whether it is possible to  realize cyons and anyons by neutral particles.  In this paper, we show that it is indeed possible.

We organize our paper as follows. In Section II, we first review the original approach of realizing a cyon by a charged planar particle and then propose a scheme to realize it by using an atom, which has a non-vanishing electric dipole  moment and a magnetic field produced by an infinitely long magnetic-charged filament. Then, in Section III, we propose a scheme to realize anyons. Compared with the case of cyon, we need one more magnetic field and a harmonic potential used to trap the atom. The differences between cyon and anyon are also analysed from the point of the conversation. The relationship between our results and previous ones is investigated from the electromagnetic duality in the last section.

\section{realize a cyon by an atom}

Traditionally, a cyon is realized by a  charged planar particle and  a perpendicular infinitely long-thin solenoid.  The  action which   describes dynamics of this charged particle  is (Latin indices run from $1$ to $2$, and summation convention is used throughout this paper)
\begin{equation}
S = \int dt \ L = \int dt \left(\half m \dot x_i ^2 - \frac{q}{c} A_i \dot x_i\right) ,  \label{laab}
\end{equation}
where $q$  is the charge that the particle carries, $c$ is speed of light in vacuum and $A_i$ is the magnetic potential produced by the infinitely long-thin solenoid. 
We choose  symmetric gauge and write the magnetic potential as
\begin{equation} \label{mab}
A_i = - \frac{\Phi}{ 2 \pi} \frac{\eps_{ij} x_j}{r^2},
\end{equation}
where $\Phi$ is the magnetic flux inside the solenoid, $\eps_{ij}$ is the Levi-Civita symbol in two-dimensional space and $r$ is the distance between the charged particle and the solenoid. It can be checked that the magnetic field corresponding to the above magnetic potential is
$B = \eps_{ij} \pa_i A_j = \Phi \del^2 (x).$
It ensures the area where the charged particle moves is magnetic-field-free.

Introduce canonical momenta with respect to  $x_i$, we get
\begin{equation}\label{pi}
p_i = \frac{\del S}{\del  \dot x_i}= m \dot x_i - \frac{q}{c} A_i.
\end{equation}
Classical Poisson brackets among canonical variables $(x_i, \ p_i)$ are
\begin{equation}
\{x_i, \ x_j \}=\{p_i, \ p_j \}=0, \quad \{x_i, \ p_j \} = \del_{ij}, \label{cpbs}
\end{equation}
which will be replaced by quantum commutators, i.e., $\{\quad, \quad \} \to \frac{1}{i \hbar} [\quad, \ \quad]$ when the canonical quantization is performed.

Considering canonical momenta (\ref{pi}), we write the canonical angular momentum  $J_c = \eps_{ij} x_i p_j$  as
\begin{equation}
J_c =  J_k - \frac{q}{c} \eps_{ij} x_i A_j \label{canangm}
\end{equation}
where
$J_k= m \eps_{ij} x_i \dot x_j$ is the kinetic angular momentum. On the other hand, the canonical angular momentum (\ref{canangm}) can also be written as $J_c = -i \hbar \pa /\pa \varphi$ with $\varphi$ being the polar angle. The single-valuedness   of the wavefunctions requires the eigenvalues of the canonical angular momentum should be quantized, i.e., ${J_c}_n = n \hbar, \ n=0, \pm 1, \pm 2, \cdots$. Substituting the explicit expression $A_i$ (\ref{mab}) into the canonical angular momentum, we find that
\begin{equation}
J_c = J_k -  \frac{q\Phi}{2 \pi c}.  \label{jm}
\end{equation}

Evidently, eigenvalues of the canonical angular momentum (\ref{jm}) are split into two parts: one is the fractional kinetic angular momentum ${J_k}_n = n \hbar + \frac{q \Phi}{2 \pi c}, \ n=0, \pm 1, \pm 2, \cdots$ which is  localized at the charged particle.  The other part $\frac{q \Phi}{2 \pi c}$ is also fractional.
The authors of Ref. \cite{cyon2}  pointed out that this part was located at the spatial infinity and was irrelevant to the studies for phenomena on a finite length. Thus, only the kinetic angular momentum is relevant to the local physical phenomena. The kinetic angular momentum with the quantum number $n=0$ divided by $\hbar$ is defined as  the spin, i.e.,
$$s= \frac{{J_k}_0}{\hbar} .$$
Therefore, the spin of a cyon is fractional, i.e. $s= \frac{q \Phi}{2 \pi \hbar c}$ \cite{wilczek10, wilczek20}.

We shall show that  a cyon can also be realized by an atom which possesses a  non-vanishing  electric dipole  moment in the background of a specific magnetic  field.
The magnetic field is produced by a long magnetic-charged filament which is parallel to the  electric dipole moment. Both the long filament and the electric dipole moment  are   perpendicular to the plane where the atom moves. This configuration  is identical to that of the He-Mckellar-Wilkens (HMW) effect, which predicted that a neutral particle with a non-vanishing electric dipole moment will acquire a topological phase if it circles around a long magnetic-charged  filament with its electric dipole moment parallel to the filament \cite{he1993, wilkens}. The  magnetic field produced by this long magnetic-charged filament is
\begin{equation}
B^{(1)}_i  = \frac{\lambda_m x_i}{2 \pi  r^2} \label{hmc1}
\end{equation}
where $\lambda_m$  is magnetic charges per unit length on the long filament. 

The Hamiltonian which describes dynamics of this  atom  is given by \cite{AC}
\begin{equation}
H = \frac{1}{2m}(p_i - \frac{d}{c^2} \eps_{ij} B^{(1)}_j )^2 \label{zha3}
\end{equation}
where $d$ is the magnitude of the electric dipole moment. Hamiltonian (\ref{zha3}) is the non-relativistic limit of  a relativistic spin-half particle which possesses a non-vanishing electric  dipole moment in the background of a magnetic field \cite{he1993, wilkens}. The action corresponding to this Hamiltonian is
\begin{equation}
S= \int dt \ L = \int dt  \ ( \half m \dot x_i ^2  + \frac{d}{c^2} \eps_{ij} \dot x_i B^{(1)}_j)  . \label{acyon}
\end{equation}

Introduce canonical momenta with respect to  $x_i$
$$p_i = \frac{\del S}{\del \dot x_i} = m \dot x_i + \frac{d}{c^2} \eps_{ij} B^{(1)}_j.$$
The canonical angular momentum is
\begin{equation}
J_c = \eps_{ij} x_i p_j = m \eps_{ij} x_i \dot x_j - \frac{d}{c^2} x_i B^{(1)}_i. \label{jc100}
\end{equation}
As usual, the eigenvalues of the canonical angular momentum are quantized, i.e., $J_{cn} = n \hbar, n=0, \pm 1, \pm 2, \cdots $. 

The canonical angular momentum (\ref{jc100}) can also be written as a summation of two parts,
\begin{equation}
J_c = J_k + J_s \label{camc1}
\end{equation}
where $J_k$ is the kinetic angular momentum and $J_s=-\frac{d}{c^2} x_i B^{(1)}_i$. Using two dimensional Dirac delta function,
$
\pa_i (\frac{x_i}{r^2}) = 2 \pi \del^{(2)} (r),
$
we can write $J_s$ as a surface term, i.e.,
\begin{equation}
J_s = -\frac{d}{2 \pi c^2} \int d^2 x \ \pa_i \big( \frac{ x_i  x_j B^{(1)}_j }{r^2}\big) .
\end{equation}
Therefore, the difference between $J_c$ and $J_k$ in the model (\ref{acyon}) is only a surface term, which takes values at the spatial infinity. However, contrary to common cases, the contribution of this surface term does not vanish.  This indicates that the eigenvalues of  canonical angular momentum are split into two fractional part, one is localized at the atom, the other is at the spatial infinity which is irrelevant to the finite length phenomena. Finally, we get the spin of the atom,
\begin{equation}
s=  \frac{\lambda_m d}{2 \pi \hbar c^2}, \label{spin}
\end{equation}
which is proportional to the magnetic charges per unit along the filament. Therefore, a cyon can also be realized by a neutral particle which possesses a non-vanishing electric dipole moment and a specific magnetic field.

\section{The realization of anyons}
In this section, we shall propose a scheme to realize an anyon.  To do that, besides the atom which possesses a non-vanishing electric dipole moment and the magnetic field (\ref{hmc1}), we need an extra magnetic field
\begin{equation}
B_i ^{(2)}= \frac{\rho_m x_i}{2} \label{b2}
\end{equation}
and a harmonic potential to trap it. The corresponding action is
\begin{equation}
S = \int dt \  ( \half m \dot x_i ^2 + \frac{d}{c^2} \eps_{ij} \dot x_i B_j ^T - \half K x_i ^2)\label{la111}
\end{equation}
where $B_i ^T = B^{(1)}_i + B_i ^{(2)}$ is the total magnetic field, $K$ is a constant which describes the intensity of the harmonic potential. To study the  rotation properties of the model (\ref{la111}), we quantize this model canonically. The canonical momenta are defined as
\begin{equation}
p_i = \frac{\del S}{\del  \dot x_i} = m \dot x_i + \frac{d}{c^2} \eps_{ij} B_j^T.
\end{equation}
After quantization, canonical variables $x_i, \ p_i$ satisfy the standard Heisenberg algebra,
\begin{equation}
[x_i, \ x_j]= [p_i, \ p_j] =0, \quad [x_i, \ p_j] = i \hbar \del_{ij}. \label{comm1}
\end{equation}

Start from the action (\ref{la111}), one can get the Hamiltonian via the Legendre transformation as
\begin{equation}
H = \frac{1}{2m}{(p_i - \frac{d}{c^2} \eps_{ij}B_j ^T)^2} + \frac{K}{2} x_i ^2, \label{hanew}
\end{equation}
which is analogous to a planar charged harmonic oscillator interacting with a  magnetic field. Similarities between neutral particles which have electric or magnetic dipole moments in the background of electromagnetic field and charged particles interacting with magnetic field was first reported in Ref. \cite{es}, in which the authors found that the Landau levels, which are spectra of a charged planar particle in the background of a uniform perpendicular magnetic field, can be simulated by a  neutral particle which possesses a permanent magnetic dipole moment interacting with a specific electric field.   Since then, the similarity between Landau levels and eigenvalues of neutral particles interacting with electromagnetic fields in various backgrounds has attracted much attention \cite{bakke1, bakke1a, bakke2, bakke3, bakke4, bakke5, bakke6, bakke7}.

Using the basic commutators (\ref{comm1}), it is easy to  prove that the canonical angular momentum
\begin{equation}
J_c = \eps_{ij} x_i p_j  \label{cam12}
\end{equation}
is not only  conserved  $[J, \ H]=0$, but also  is the generator of rotation transformation,
$[J, \ x_i] = i \hbar \eps_{ij} x_j, \quad [J, \ p_i] = i \hbar \eps_{ij} p_j$. Obviously, the eigenvalues of the canonical angular momentum can only  be integer.

Now, we investigate rotation properties of the reduced model which is the limit of  cooling down the atom to the negligibly small kinetic energy. This kind of  limit was  considered in the Chern-Simons quantum mechanics \cite{jackiw}.  Mathematically, this limit amounts to set the kinetic energy term in the action (\ref{la111}) to zero. The action of the reduced model can be easily obtained from the full one (\ref{la111}) by neglecting kinetic energy as
\begin{equation}
S_r = \int dt \  (\frac{d}{c^2} \eps_{ij} \dot x_i B_j ^T - \half K x_i ^2). \label{ra}
\end{equation}
Introduce canonical momenta with respect to $x_i$
\begin{equation}
p_i = \frac{\del S}{\del \dot x_i} = \frac{d}{c^2} \eps_{ij} B_j ^T.
\end{equation}{} This results in  primary constraints,
\begin{equation}
\phi_i    = p_i - \frac{d}{c^2} \eps_{ij} B_j ^T \approx 0.  \label{pcs}
\end{equation}
In  terminology of Dirac, the symbol `$\approx$' is weak equivalence which  means equivalent only on the constraint hypersurface. The existence of constraints indicates that there are redundant degrees of freedom. The Poisson brackets between primary constraints are
\begin{equation}
\{ \phi_i   , \ \phi_j    \}= - \frac{d \rho_m}{c^2} \eps_{ij}.
\end{equation}
According to the classification of constraints,  primary constraints (\ref{pcs}) belong to second class and can be used to get rid of redundant degrees of freedom.

The canonical angular momentum of the reduced model (\ref{ra}) is also defined as $J= \eps_{ij} x_i p_j$. However, since  there are redundant degrees of freedom, $x_i$ and $p_i$ are not independent. The redundant degrees of freedom can be eliminated by substituting constraints (\ref{pcs}). In terms of independent variables, we write  the canonical angular momentum  as
\begin{equation}
J_r =\eps_{ij} x_i p_j = - \frac{d}{2c^2} \left(\rho_m x_i ^2 + \frac{\lambda_m}{\pi}\right). \label{camrm} 
\end{equation}
Because of the primary constraints (\ref{pcs}), the standard  Poisson brackets  (\ref{cpbs}) among canonical variables $(x_i, \ p_i)$ should be replaced by Dirac brackets
\begin{equation}
\{x_i, \ x_j \}_D = \{x_i , \ x_j \} -\{ x_i, \phi_m    \} \{\phi_m   , \ \phi_n    \} ^{-1} \{\phi_n   , \ x_j \}
\end{equation}
After a direct calculation, we get
$\{x_i, \ x_j \}_D = \frac{c^2}{d \rho_m} \eps_{ij}, $
from which we have
\begin{equation}
[x_i, \ x_j] = \frac{i \hbar c^2}{d \rho_m} \eps_{ij}.  \label{cr111}
\end{equation}{}It is easy to find that the canonical angular momentum (\ref{camrm}) is analogous to a one-dimensional harmonic oscillator. We can get  eigenvalues of $J_r$  directly
\begin{equation}
J_{rn} = - (n+ \half ) \hbar - \frac{  \lambda_m d}{2 \pi c^2}, \ \ n=0,1,2, \cdots. \label{fcam}
\end{equation}

Obviously, apart from a normal part, the eigenvalues of the canonical angular momentum contain a fractional part which is equivalent to the one we studied in previous section. It is showed that  eigenvalues of the canonical angular momentum  can take fractional values.

There is a great difference between results (\ref{spin}) and (\ref{fcam}).  The canonical angular momenta (\ref{cam12}) and  (\ref{camrm}) are always conserved. However, the kinetic angular momentum is not.
It  will be more transparent  if we consider the case of  varied  line density of the magnetic charges, i.e., $\lambda_m= \lambda_m(t)$. Both  (\ref{cam12}) and  (\ref{camrm}) are still conserved since they are all  Noether charges of the planar rotation transformation ($\del \varphi$ is an infinitely small angle)
\begin{equation}
x_i \to x_i ^\prime = x_i +  \del\varphi \eps_{ij} x_j  \label{prt}
\end{equation}
regardless  the line density $\lambda_m$ is time-dependent or not. It means that 
$\dot J_c =0$.  
As a result, we get $\dot s =   \frac{d\dot \lambda_m (t)}{2 \pi \hbar c^2 } \neq 0$. Therefore, the kinetic angular momentum and the  spin of a cyon are not conserved in the case of varied line density of the magnetic charges.

\section{Conclusions and further discussions}

In this paper, we propose to realize cyons and anyons by using atoms which have non-vanishing electric dipole moments. The results presented here can also be understood from the electromagnetic duality point of view.

Historically, after the discovery of AB effect \cite{AB}, Aharonov and Casher predicted that a neutral particle with a non-vanishing magnetic dipole moment would acquire a  topological phase if it moves around a uniformly charged infinitely
long filament with its direction paralleling to the filament. It is the Aharonov-Casher (AC) effect \cite{AC, EAC, Boyer, APV, Goldhaber}. It is generally accepted that the AB effect is dual to the  AC effect in the sense that the solenoid in the AB effect can be thought of  a linear array of magnetic dipoles
and exchanges the magnetic dipoles with the electric charges  \cite{AC, sv}. Therefore, for the AC effect one has a line of electric charges and a particle with a magnetic dipole moment moving around this line. 

The Hamiltonian describing the interaction between a neutral particle with a non-vanishing magnetic dipole moment and  electric fields is
\begin{equation}
H = \frac{1}{2m} (  p_i + \frac{\mu}{c^2} \eps_{ij} E_j)^2
\label{hac}
\end{equation}
where $\mu$ is the magnitude of the magnetic dipole moment. It is the non-relativistic limit of a relativistic spin half neutral particle with a non-vanishing magnetic dipole moment interacting with electric fields \cite{AC}.

In  \cite{jzd}, the authors realized the fractional angular momentum from the above Hamiltonian by applying two electric fields  $E_i ^T =  E_i^{(1)} +  E_i^{(2)}$ and confining it by a harmonic trap potential.
Two electric fields  are
\begin{eqnarray}
E_i^{(1)} &=& \frac{\lambda_e x_i}{2 \pi \eps_0 r^2}, \label{efs1} \\
E_i^{(2)} &=& \frac{\rho_e x_i}{2 \eps_0},  \label{efs2}
\end{eqnarray}
where $\lambda_e$ and $\rho_e$ are the charges per unit length on the long filament and the charge density respectively, $\eps_0$ is the dielectric constant. After taking the limit of cooling down the kinetic energy of the atom to its lowest level, one gets the canonical angular momentum as \cite{jzd}
\begin{equation}
J = \frac{\mu}{2 c^2 \eps_0} \left(\rho_e x_i ^2+ \frac{\lambda_e}{\pi}\right) \label{camp}
\end{equation}
where $x_i$ are coordinates of the atom on the plane which is perpendicular to the magnetic dipole moment. They satisfy the commutation relations
\begin{equation}
[x_i, \ x_j] = - \frac{i \hbar \eps_0 c^2 }{\mu \rho_e}. \label{commp}
\end{equation}
Obviously, the eigenvalues of the canonical angular momentum can be read directly from  (\ref{camp}) and  commutation relation (\ref{commp}) as \cite{jzd}
\begin{equation}
J_n = (n + \half) \hbar + \frac{\mu \lambda_e}{2 \pi c^2 \eps_0}.
\end{equation}

However, it is argued that it is the HMW effect  \cite{he1993} rather than the AB effect that is dual to the AC effect.
Because there is a natural correspondence between the AC and HMW effects: the magnetic dipole moment in the AC effect corresponds to the electric dipole moment in the HMW effect, the electric charges in the AC effect corresponds to the magnetic charges in the HMW effect.

The duality between the AC effect and the HMW effect can be expressed exactly as
\begin{equation}
\left\{\begin{array}{c} \mathbf E \\ d \end{array} \right \} \to \frac{1}{\sqrt {\eps_0 \mu_0}} \left\{\begin{array}{c} \mathbf B \\ \mu \end{array} \right \}, \quad
\left\{\begin{array}{c} \lambda_e \\ \rho_e   \end{array} \right \} \to  \sqrt{\frac{\eps_0}{\mu_0}} \left\{\begin{array}{c} \lambda_m \\ \rho_m  \end{array} \right \} \label{dr1}
\end{equation}
and
\begin{equation}
\left\{\begin{array}{c} \mathbf B \\ \mu \end{array} \right \} \to - {\sqrt {\eps_0 \mu_0}} \left\{\begin{array}{c} \mathbf E \\ d \end{array} \right \}, \quad
\left\{\begin{array}{c} \lambda_m \\ \rho_m   \end{array} \right \} \to  -\sqrt{\frac{\mu_0}{\eps_0}} \left\{\begin{array}{c} \lambda_e \\ \rho_e \end{array} \right \} \label{dr2}
\end{equation}
where the minus arises from the asymmetric nature of the electromagnetic duality.

It is easy to check that the AC effect is dual to  the HMW effect through (\ref{dr1}) and the HMW effect is dual to the AC effect through (\ref{dr2}). Furthermore, the Maxwell equations are also invariant under the dualities (\ref{dr1}), (\ref{dr2}) and
\begin{equation}
J_e \to \sqrt{\frac{\eps_0}{\mu_0}} J_m, \ J_m \to - \sqrt{\frac{\mu_0}{\eps_0}} J_e
\end{equation}
with  $J_e \ (J_m) $ being the electric  (magnetic) current density provided the magnetic charges are presented \cite{jackson}. Interestingly, the authors of \cite{dwf} predict that  there is a  new topological effect which is dual to the AB effect according to the duality relations (\ref{dr1}) and (\ref{dr2}).

These duality relations are also applied in \cite{braz}, in which the authors find that  energy spectra of an electric dipole moment in the background of the magnetic field  (\ref{b2}) are dual to spectra of a magnetic dipole moment interacting with the electric field (\ref{efs2}) through  (\ref{dr1}, \ref{dr2})  \cite{es}. 

The  dualities (\ref{dr1}) and (\ref{dr2}) not only hold  in the Hamiltonians (\ref{hanew}) and  (\ref{hac}), but also in the canonical angular momentum (\ref{camrm}) and (\ref{camp}) as well as the commutation relations between $x_i$ in the reduced models (\ref{cr111}) and  (\ref{commp}). Therefore, the studies of the present paper can be regarded as the electromagnetic duality of the \cite{jzd} from the point of view  (\ref{dr1}) and (\ref{dr2}). It may be curious that eigenvalues of the angular momentum (\ref{fcam}) take negative values. The minus sign in  eigenvalues of the angular momentum (\ref{fcam}) originates from (\ref{camrm}), which is the direct result of the electromagnetic dualities (\ref{dr1}) and (\ref{dr2}). It can also be understood classically since  the atom can only stay in the stable equilibrium  provided  the direction  of the classical angular momentum is inverse to the electric dipole moment.

\section*{Acknowledgements}

This work is supported by NSFC with Grant No. 11465006 and partially supported by 20190234-SIP-IPN and the CONACyT under grant No. 288856-CB-2016.


\begin{thebibliography}{99}

\bibitem{anyons1}
A. Lerda, Anyons, Quantum Mechanics of Particles with Fractional Statistics, Lecture Notes in Physics, New Series m: Monographs,  Springer-Verlag,
Berlin, 1992.


\bibitem{anyons2}

A. Khare, Fractional Statistics and Quantum Theory, World Scientific, Singapore, 1997.


\bibitem{hall}
T. Chakraborty and P. Pietil$\ddot {\rm a}$inen, The Quantum Hall Effect, Fractional and Integral, Second Ed, Springer-Verlag, Berlin, Heidelberg, 1995.


\bibitem{htc}
F. Wilczek, Fractional Statistics and Anyon Superconductivity, World Scientific, Singapore, 1990.


\bibitem{cs1}
S. Deser, R. Jackiw, and S. Templeton, Ann. Phys. (N.Y.) 140 (1982) 372.

\bibitem{cs2}
S. Zhang, T. Hanson, and S. Kivelson, Phys. Rev. Lett. 62 (1989) 82.

\bibitem{cs3}
R. Jackiw and So-Young Pi, Phys. Rev. D 44 (1991) 2524.

\bibitem{forte}
S. Forte, Rev. Mod. Phys 64 (1992) 193.

\bibitem{zhang1}
Y. Zhang, G. J. Sreejith, N. D. Gemelke and J. K. Jain, Phys. Rev. Lett 113 (2014) 160404.

\bibitem{zhang2}
Y. Zhang, G. J. Sreejith, and J. K. Jain, Phys. Rev. B 92 (2015) 075116.

\bibitem{zhang3}
D. Lundholm and N. Rougerie, Phys. Rev. Lett. 116 (2016) 170401.

\bibitem{zhang}
J. Z. Zhang, Phys. Lett. B, 670 (2008) 205.

\bibitem{cyon1}
A. S. Goldhaber, Phys. Rev. Lett. 49 (1982) 905.

\bibitem{cyon2}
A. S. Goldhaber and R. Mackenzie, Phys. Lett. B 214 (1988) 471.

\bibitem{wilczek10}
F. Wilczek, Phys. Rev. Lett. 48 (1982) 114.

\bibitem{wilczek20}
F. Wilczek, Phys. Rev. Lett. 49 (1982) 957.

\bibitem{he1993}
X. G. He and B. H. J. Mckellar, Phys. Rev. A. 47 (1993) 3424.

\bibitem{wilkens}
M. Wilkens, Phys. Rev. Lett. 72 (1994) 5.

\bibitem{es}
M. Ericsson and E. Sj\"{o}qvist, Phys. Rev. A 65 (2001) 013607.


\bibitem{bakke1}
K. Bakke and C. Furtado, Phys. Rev. A 80 (2009) 032106.

\bibitem{bakke1a}
K. Bakke, L. R. Ribeiro, C. Furtado and J. R. Nascimento, Phys. Rev. D 79 (2009) 024008.

\bibitem{bakke2}
K. Bakke, L. R. Ribeiro, C. Furtado, Cent. Eur. J. Phys. 8  (2010) 893.

\bibitem{bakke3}
A. B. Oliveira, K. Bakke, Proc. R. Soc. A  472 (2016) 20150858 .

\bibitem{bakke4}
A. B. Oliveira, K. Bakke, Int. J. Mod. Phys. A  31 (2016) 1650019.

\bibitem{bakke5}
A. B. Oliveira, K. Bakke, Eur. Phys. J. Plus 131 (2016) 266.

\bibitem{bakke6}
A. B. Oliveira, K. Bakke, Ann. Phys. 365 (2016) 66.

\bibitem{bakke7}
A. B. Oliveira, K. Bakke, R. Soc. Open Sci. 4 (2017) 170541.


\bibitem{jackiw}
G. V. Dunne, R. Jackiw, and C. A. Trugenberger, Phys. Rev. D 41 (1990) 661.

\bibitem{AB}
Y. Aharonov and D. Bohm, Phys. Rev. 115 (1959) 485.

\bibitem{AC}
Y. Aharonov and A. Casher, Phys. Rev. Lett. 53 (1984) 319.


\bibitem{Boyer}
T. H. Boyer, Phys. Rev. A. 36 (1987) 5083.

\bibitem{APV}
Y. Aharonov, P. Pearle and L. Vaidman, Phys. Rev. A. 37 (1988) 4052.


\bibitem{Goldhaber}
A. S. Goldhaber, Phys. Rev. Lett. 62 (1989) 482.

\bibitem{EAC}
A. Cimmino, G. I. Opat, A. G. Klein, H. Kaiser, S. A. Werner, M. Arif, and R. Clothier,  Phys. Rev. Lett. 63 (1989) 380.

\bibitem{sv}
D. Singleton and  E. Vagenasc, Phys. Lett. B. 723 (2013) 241.

\bibitem{jzd}
J. Jing, Q. Y. Zhang, Q. Wang, Z. W. Long and S. H. Dong, Eur. Phys. J. C 79 (2019) 1.

\bibitem{jackson}
J. D. Jackson, Classical Electrodynamics, 3rd Edition, Published by John Wiley $\&$ Sons, Inc.

\bibitem{dwf}
J. P. Dowling, C. P. Williams and J. D. Franson, Phys. Rev. Lett. 83 (1999) 2486.

\bibitem{braz}
L. R. Ribeiro, C. Furtado and  J. R. Nascimento, Phys. Lett. A 348 (2006) 135.


\end{thebibliography}
\end{document}